# High-Q, size-independent, and reconfigurable optical antennas via zero-index material dispersion engineering

Prasad P. Iyer[1,4*], Mihir Pendharkar[1,5], Anchal Agrawal[1,3], Humberto Foronda[1,3], Mike Iza[2,3], Umesh K. Mishra[1,3], Shuji Nakamura[1,2,3], Steven DenBaars[1,2,3], Stacia Keller[2,3], Chris Palmstrøm[2] and Jon A. Schuller[1*]

[1.]Electrical and Computer Engineering Department, University of California Santa Barbara, CA. [2.]Material Science and Engineering Department, University of California Santa Barbara, CA. [3.]Solid State Lighting and Energy Center, University of California Santa Barbara, CA [4.]Center for Integrated Nanotechnologies, Sandia National Lab, Albuquerque, NM. [5]Materials Science and Engineering department, Stanford University, Stanford, CA.
*ppadma@sandia.gov, jonschuller@ece.uscb.edu

**Abstract**

Enhancing light-matter interactions at the nanoscale is foundational to nanophotonics, with epsilon-near-zero (ENZ) materials demonstrating significant potential. High-quality (Q) factor resonances maximizing these interactions are typically realized in photonic crystals requiring sub-50 nm precision nanofabrication over large areas, limiting scalability and increasing complexity. Mie resonances offer an alternative but are constrained by low Q-factors due to the scarcity of high refractive index materials, necessitating large refractive index changes for effective resonance switching and limiting dynamic reconfigurability. We overcome these limitations by embedding Mie resonators within ENZ media, thereby enhancing Q-factors, mitigating geometric dispersion and fabrication challenges, and maximizing optical reconfigurability. We introduce three resonator-ENZ configurations: voids in AlN, Ge in $SiO_2$, and intrinsic InSb in doped InSb—spanning from low-loss phononic to lossy plasmonic ENZ modes. Using novel epitaxial regrowth techniques, we achieve significant Q-factor improvements over non-embedded resonators. Notably, an air-based Mie resonator embedded in AlN supports resonant Q-factors exceeding 100, with negligible geometric dispersion across sizes from 800 nm to 2,800 nm. Additionally, we demonstrate dynamic reconfigurability of intrinsic InSb resonators by thermally tuning the ENZ wavelength over a 2 μm range in the mid-infrared (11–16 μm) wavelength regime. These results showcase the potential of Mie reonators embedded in ENZ media for high-fidelity sensors, thermal emitters, and reconfigurable metasurfaces, bridging theoretical predictions with practical applications, advancing the development of dynamic, high-Q optical devices.

**Introduction**

Enhancing light-matter interactions at the nanoscale forms a foundational pillar of nanophotonics, with zero refractive index (ZI) or epsilon-near-zero (ENZ) resonant systems emerging as a powerful method to maximize these interactions[1–7]. ENZ materials offer tremendous potential across applications such as redirecting thermal emission[8–10], super radiance[11,12], lensing[13], photonic doping[14], squeezing[15–17], tunnelling light[18] and enchanced non-linear[19–22] interactions. Maximizing light-matter interactions using metasurface resonances minimizes the magnitude of reconfigurability necessary for re-structuring light [23–30], increases non-linear phenomena [31,32], and enables novel optoelectronic devices [33–44].

Recently, optical resonances within high-index dielectric metasurfaces and photonic crystal slabs, consisting of sub-wavelength arrays of low-loss optical resonators, have emerged as prime candidates for realizing high-Q (quality factor) resonances for designing spatio-temporal responses of light-matter interactions at the nanoscale. However, while photonic crystal lattice-based resonances have been successful in realizing high-Q (approaching infinity) quasi-bound-state-in-continuum resonances, they require translational symmetry in the lattice and necessitate extremely precise (< 50nm-scale) fabrication over large areas (>500μm$^2$) [33,45–47]. This requirement limits large-area scalable fabrication of spatial phase profiles necessary for meta-optical components. Low-loss Mie resonators, on the other hand, are highly efficient building blocks of metasurfaces (enabling complete reconfigurable control over the amplitude[48], phase[49] and polarization[50] of light) due to their ability to radiatively couple to light and consequently suffer from low Q. Although high-refractive index materials enhance the Q of the resonances, there is a fundamental limit to the maximum refractive index a material can achieve within the optical frequency domain [51–53]. Furthermore, the amount of refractive index tuning necessary to switch the resonance by its linewidth (full-width half max) scales inversely with the Q. As the resonant quality factor increases, the dispersion of the resonances with respect to geometrical parameters also increases dramatically, reducing the available fabrication tolerances to realize such high-Q resonances. Specifically, the magnitude of achievble active tuning of Mie resonantors[24,28–30,54] limits its applicability in realizing actively reconfigurable metasurfaces.

To address these limitations, researchers have proposed embedding Mie resonators within ENZ materials[55–60]. This approach can both enhance Q-factors and improve fabrication tolerances by obviating geometric dispersion issues. Despite promising theory results, conclusive experimental validation of these phenomena are lacking, largely because existing eperimental studies are based on Mie resonators sitting *atop* ENZ media. Here we demonstrate novel growth and fabrication methods for embedding Mie resonators within ENZ media and experimentally demonstrate greater than an order-of-magnitude increases in the Q-factor of fundamental scattering resonances and size-independent pinning of resonances to the ENZ wavelength. Furthermore, through dynamic refractive index modulation of the ENZ material, we demonstrate size-independent tuning of the embedded Mie resonances.

Traditional semiconductor and dielectric optical antennas comprise high refractive index resonators ($n_{res}$) embedded within a low-index background medium ($n_{bkg}$). Mie resonances correspond to eigensolutions of electromagnetic scattering equations, and follow the relationship $\frac{n_{res} \cdot r_{res}}{n_{bkg} \lambda_{res}} = C_{Mie}$, where $C_{Mie}$ is the resonant size parameter (Figure 1A)[61]. The quality factor of these resonances within the individual scattering antennas[62,63] is proportional to $\left(\frac{n_{res}}{n_{bkg}}\right)^m$, $where\ m > 1$. In the scenario where, $n_{bkg}$ → 0 at $\lambda_{ENZ}$, a new regime of antenna operation has been theoretically predicted due to the pole in the permittivity of the background media (Figure 1B): (i) the geometric dispersion relationship collapses; at the ENZ wavelength all resonator sizes satisfy the resonance condition, and (ii) the Q-factor diverges to infinity in the ideal scenario of no absorption losses. Here, we experimentally demonstrate novel embedding approaches and phenomena within three distinct material systems: a) Voids in AlN, b) Ge in SiO2, and c) intrinsic InSb in doped InSb.

Experimentally achieving theoretically-predicted ENZ properties such as extremely high quality (Q) factors, size-independent resonances, or exotic bound states[34] has proven cumbersome due to absorption losses and the fabrication challenges associated with embedding resonators within ENZ media[6,20,64,65]. In fact, nearly all existing experimental demonstrations of these phenomena are based on Mie resonators sitting *atop* ENZ media. In the optical frequency regime, ENZ properties derive from either optical phonons (polaritonic) or free carriers

(plasmonic). In the first case, compound oxides (e.g., $SiO_2$, $Al_2O_3$), nitrides (e.g., $Si_3N_4$, AlN, GaN), and carbides (e.g., SiC) exhibit ENZ behavor at the longitudinal optical (LO) phonon frequency, where the real part ($\epsilon_r$) of the dielectric constant passes through zero.[66,67] Relative to plasmonic ENZ, the absorption (determined by the imaginary part of the dielectric constant, $\epsilon_i$ at the LO phonon frequency) is quite low,[68,69] making these material systems ideal test beds for studying properties of ENZ-embedded Mie resonators .[70]

To illustrate this, we first analyze the thin-film reflection of a 2μm AlN layer grown on SiC via metal-organic chemical vapor deposition (MOCVD). By fitting the reflection spectra using a transfer matrix approach, we extract the complex dielectric permittivity of the thin film (Supplementary information S1, Figure S1A and S1B). The imaginary part of the permittivity $\epsilon_i$ at the ENZ wavelength is 0.022±5e-3, the lowest value reported for an ENZ material. Here, we developed a novel metal-organic chemical vapor deposition (MOCVD) regrowth process for fabricating air-void nanowires encapsulated in AlN through a scalable, three-step approach. This process involves the initial MOCVD growth of AlN, nanofabrication of nanowire trenches, and a two-step epitaxial regrowth of GaN nanowires and AlN thin films. GaN nanowires are selectively grown within AlN trenches, which are etched using a $SiO_2$ hard mask that also serves to isolate GaN growth on AlN surfaces. The $SiO_2$ hard mask is subsequently removed with a dilute HF acid wet etch, and the sample is cleaned in preparation for the second AlN regrowth phase. The two-step AlN regrowth process begins with a low-temperature deposition to form a thin nucleation layer, followed by high-temperature AlN growth. This low-temperature growth establishes a thin AlN layer on the GaN nanoantennas, while the subsequent high-temperature step promotes the evaporation of the GaN nanoantennas, resulting in the formation of air-void antennas. In addition to achieving record-low losses, the incorporation of non-dispersive, low-index air voids offers valuable insights into the fundamental limits of resonators embedded in ENZ (epsilon-near-zero) media. The air voids within AlN serve as exemplars of low-index resonators embedded in a phononic ENZ medium. To contrast this with high-index Mie resonators, we embedded Ge resonators (n $\approx$ 4) in $SiO_2$, which serves as the ENZ medium. Although $SiO_2$ is inherently more lossy than AlN, its longitudinal optical (LO) phonon ENZ point at 7.9 μm has an imaginary permittivity, $\epsilon_i$ of 0.5 (see Supplementary Information S1, Figure S1C and S1D), which, while more absorptive than the corresponding LO-phonon mode of AlN, remains significantly less absorptive than plasmonic ENZ materials. For this configuration, we employ electron-beam deposition to place Ge nanowires within etched $SiO_2$ trenches, followed by encapsulation in $SiO_2$ via a plasma-enhanced CVD process, ensuring that the Ge antennas are entirely embedded within the glass matrix.

One limitation of polaritonic ENZ media is the lack of tunability; ENZ wavelengths are determined by intrinsic phonon properites. In contrast, plasmonic ENZ materials are formed through the introduction of Drude free-carriers, which can be achieved in semiconductors via either static (doping) or dynamic (optical or electrical pumping) processes[71–73]. These free-carriers not only reduce the real part of the refractive index but also increases the imaginary part, resulting in increased absorption of light within the media. Intrinsic InSb is a high-index semiconductor whose mid-infrared refractive index can be changed statically or dynamically through free carrier refraction. InSb is an ideal ENZ semiconductor because of its low band-minimum electron effective mass ($m_e$ = $0.014m_o$) and high electron mobility ($\mu_e$) [28,74,75]. This allows InSb to form an ENZ medium in the mid-infrared wavelength regime with low doping concentrations (low $m_e$), and minimal absorption losses (high $\mu_e$). Nonetheless, measured values of $\varepsilon_i$ ~1.5 at ENZ condition for n-InSb (Si: $5\times10^{18}cm^{-3}$) thin-films grown on i-GaSb (Supplementary information Section S1, Figure S1E and S1F) are higher than the polaritonic systems described previously. Here, we fabricate intrinsic high-index InSb wires embedded within a doped ENZ InSb medium using a novel surface preparation method (including alternate wet surface oxidation with $HNO_3$ and removal with buffered HF followed by an atomic hydrogen clean clean in ultra high vacuum), that enables single-crystal epitaxial regrowth of InSb on a dry-etched InSb surface by Molecular Beam Epitaxy (MBE). In this system, we demonstrate dynamic reconfigurability via thermal control of the ENZ wavelength, showcasing the potential for real-time tuning of optical properties.

The three different ENZ media enable three resonant embedded nanowire systems with ENZ wavelengths of 7.9 µm (Ge in $SiO_2$), 11.2 µm (voids in AlN), and 13-15 µm (i-InSb in n-InSb).

Using an FTIR microscope we measure single nanowire (diameters between 800-2800 nm) resonances via differential reflectivity (D.R.), in order to demonstrate size-independent Mie resonances and Q-factor enhancements. Size-independent Mie resonances ultimately derive from the pole in the dielectric permittivity of the embedding ENZ medium, which forces the resonance condition to be independent of the size or refractive index of the resonator. The Eigensolution for all Mie resonances is determined by equation 1, where $n_{res}$ ($n_{bkg}$) is the refractive index of the resonator (background), r is the size of the resonator (eg. Radius of the nano-wire), $\lambda_{res}$ is the resonant wavelength and $C_{Mie}$ represents a resonant size parameter (for example in InSb Mie resonators: $C_{Mie}$= 0.64 for the electric dipole and $C_{Mie}$= 0.48 for the magnetic dipole resonances). In the case when the $n_{bkg}$ goes to zero at the ENZ wavelengths, the size dispersion of the resonance condition boils down to a trivial solution independent of the size and index of the resonator.

$$\frac{n_{res}}{n_{bkg}} \cdot \frac{r}{\lambda_{res}} = C_{Mie}: n_{bkg} \to 0 \xrightarrow[\lambda_{res} \to \lambda_{ENZ}]{} \begin{cases} r \to 0 \\ \delta r \to 0 \end{cases}$$

Demonstrations of pinning are demonstrated and discussed individually for each material system, followed by demonstrations of Q-factor enhancement and tunability.

**Air-Voids in AlN:** Traditional void resonators comprise air voids embedded within *higher* index media. Such void resonances can be nearly lossless—albeit with very low-Q (2-3)—and have been studied for solar cell architecture[76,77] and microfluidics[78] applications. Here, we demonstrate a new kind of void resonator where the embedding medium has a refractive index (close to zero) lower than that of the void. In essence, we increase the *effective* refractive index of vacuum above one by embedding it in a ZI media. The unique dielectric boundary condition slows[79] the light down and forces the electric field to concentrate within the void. Electromagnetic simulations reveal [76,78,79] a fundamental TE- (TM-) polarized electric (magnetic) dipole resonance at the AlN ENZ wavelength (11.2μm). These resonances are associated with a derivative-like lineshape in the experimental DR spectra (orange band in Fig 2B). The lack of dispersion w.r.t the void size (Fig. 2B) demonstrates that the ENZ resonances (11.2μm) are completely size independent. Similar resonance pinning of plasmonic resonators sitting atop ENZ substrates was previously observed, but limited to a small size regime[20,80]. Here, the embedded nature of the voids produces both TM- and TE-polarized (Fig 2B) resonance pinning across a large variation in resonator widths. The large design window of these resonators alleviate the common tolerances challenges experiences during nanofabrication of large area metasurfaces.

**Ge in SiO$_2$:** In comparison with air voids in AlN, Ge nanowires embedded within SiO$_2$ (methods) provide insight into the behavior of high index resonators embedded in ENZ media. The SiO$_2$ ENZ wavelength ($\varepsilon_i$ ~ 0.5) is 7.9μm. Similar to voids in AlN, both TE (electric dipole) and TM (magnetic dipole) Mie resonances are supported at the ENZ wavelength (Fig. 2C). Here, however, resonances are associated with minima in the differential reflectivity (Fig 2D). Like air voids, the resonances are pinned to the ENZ wavelength and exhibit nearly zero size dispersion. The magnitude of the TE absorbtion is much higher than the TM absorption as expected since an ideal lossless ENZ media behaves like a perfect magnetic conductor for the TM polarized incident wave.[81,82]

**InSb resonators:** Unlike previous cases, the regrowth here is more conformal with a doped-InSb nanowire located atop the embedded i-InSb resonator (Fig 2E). TM polarized light excites resonances in the embedded high-index nanowire Mie resonator whereas TE polarized light excites plasmonic resonance overlying regrown n-InSb nanowire. The geometric dispersion of the TM resonances demonstrates similar behavior to similar resonances for Ge antennas in SiO$_2$. As the size of the resonator becomes smaller, the resonance wavelength shifts towards the ENZ wavelength, ie. arbitrarily small sized antennas can form resonances with large scattering cross section at the ENZ wavelength. Relative to the polaritonic systems, there is more size dispersion due to the lossier nature of the embedding medium and the associated reduction in intrinsic material permittivity dispersion. Nonetheless, the resonance wavelength shifts by only 1.78% despite a 300% increase in resonator width, demonstrating that the system exhibits pinning behavior. The TE resonance for the same resonator (the conformal fin formed during re-growth of n-InSb) serves as a remarkable control since it is a plasmonic resonator which disperses with width of the resonator by over 4X the TM resonance enabling us to directly compare and quantify the pinning effect of resonators embedded in ENZ media.

**Q-factor comparison:** Comparing the three different ENZ media with different absorption coefficients at the ENZ wavelenghts, ($\varepsilon_i$ ~ $10^{-2}$, 0.5 and 1.5 for AlN, SiO$_2$ and n-InSb respectively), we can directly quanitfy the impact of the losses for the embedded resonators. The lineshapes of the resonances in the differential reflectivity plots potray the effect of decreasing losses of the phonon modes interacting with the loss-less dielectric Mie reosnators. Specifically, D.R resonant spectra transforms from derivative line-shape in the low-loss AlN phonon case to a classical lorentzian lineshape for more absorbing ENZ modes[83,84]. Measured Q-factors for the ENZ-coupled TM resonances are shown in Figure 3A. The Q-factor depends pricncipally on the lossiness of the ENZ-embedding medium. The low absorption ($\varepsilon_i$ ~ $10^{-2}$) at the LO phonon of AlN enables us to acheive a high-Q factor (95). This enhancement is particularly notable when comparing to the "un-embedded" alternative; an air-void sitting atop the substrate supports no resonance at all. Although direct experimental comparisons cannot be made for Ge-in-SiO$_2$, measured Q values between 45-50 are approximately an order-of-magnitude larger than calculations for Ge nanowires in air (Q=5) and similar values reported in the literature.[24,30,85,86] For the InSb system, we directly compare Q-factors for i-InSb sitting *atop* the substrate vs. embeded *within* the substrate; embedding the resonators increases the Q-factor by 400% (Fig. 3). [85][61]

**Thermal Tunability of InSb resonators:** Unlike phononic ENZ resonators[87], the ENZ wavelength of the n-InSb can be modulated by a large magnitude due to the variation the electron effective mass variation as demonstrated previously[28]. This enables tuning of the ENZ-coupled resonances based on the refractive index modulation of the doped InSb medium. The n-InSb thin-film reflectance exhibits continuous modulation of the absolute reflectance by over 60% with 493 K temperature swing. (Fig 4A) This large variation in the amplitude of the reflected light is comparable to insulator to metal switching materials but this semicondcutor material system provides continuous and homogenous reconfigurability of the refractive index as a function of temperature.[85,86] The ENZ wavelength and the imaginary part of the perimittivity at the ENZ wavelength can be extracted by fitting the reflection curves with a Fabry-Perot transfer matrix approach. We show that the ENZ wavelength red shifts from 13.1μm to 15.0μm while $\varepsilon_i$ at the ENZ wavelength increases from 1.35 to 1.55 when the temperature of the substrate is increased from 80K to 573K. (Fig 4B) The large magnitude of the change in the refractive index of the material makes it an ideal choice to tune low-Q Mie and plasmonic resonances. We demonstrate dynamic thermal tuning of an TM ENZ-coupled nanowire resonance in Figure 4C. The increase in the Q-factor due to embedding along with the large refractive index change enables us to tune the resonance by 1.7 times the line-width of the resonance.

**Conclusion**

In summary, we have experimentally demonstrated first examples of Mie resonators embedded within three different ENZ media. We developed novel fabrication and epitaxial re-growth strategies to embedd high-index and low-index resonators within ENZ media over range semiconductor material systems. We demonstrate the potential of ENZ[14,88,89] materials to enable deeply subwavelength optical resonances that transcend the size and refractive index limitations of conventional optical antennas. All systems demonstrate significant Q-factor enhancements—up to nearly 100-fold—and reduction in size dispersion. Comparisons across the different material systems confirm that absorption at the ENZ wavelength primarily limits the maxium possible Q-factors achievable for high-index dielectric resonators. Additionally, we address the challenge of dynamic reconfigurability by demonstrating thermal control of the ENZ wavelength in InSb, showcasing the potential for real-time tuning of optical properties. In InSb, the combination of Q-factor enhancement and large thermal reconfigurability enables resonance tuning by one line-width with ~ 200 K temperature shift. These results bridge the gap between theoretical predictions and practically viable applications of ENZ resonators to high fidelity sensors, thermal emitters and reconfigurable metasurfaces.

**Methods**

**Voids in AlN :** We have developed a novel fabrication process to naturally form void nanowires of controlled sizes within the thin film of single crystal AlN (2µm) on SiC substrate without any substrate removal or bonding. We expitaxially grow AlN (using MOCVD) on SiC substrate which has its LO phonon at 10.32µm and thus behaves like a reflecting metal at 11.2µm where we form our ENZ resonators. Intially, we etch nanowire pits of different widths into AlN thin film on SiC using a 500nm $SiO_2$ hard mask of required dimensions defined using photolithography. The etched sample with the oxide hard mask is prepared for GaN regrowth in the MOCVD chamber. Due to large lattice mismatch and the amorphous nature of the $SiO_2$, GaN selectively regrows within the etched pits to fill up volume of AlN previously etched. The GaN was grown at 1.2A/s under 600 torr gas pressure at 1030°C temperature without any initial nucleation layer to ensure growth from the sidewalls. At this point, the oxide hard mask is removed using a buffered HF (BHF) wet etch procedure. Subsequently, a blanket AlN regrowth is done on this sample. This is done with a nucleation layer of AlN at the GaN growth temperature (1030°C) and then the temperature is increased to AlN growth temperature (1250°C) at 100 torr pressure. Such a growth procedure intially forms a conformal nucleation layer of AlN on the entire film while the second stage of the growth process at the higher temperature forces the GaN to desorb from within the etch resonator pits. This forms the voids while AlN regrows conformally on top. We have thus developed a novel fabrication process to naturally form void nanowires of controlled sizes within the thin film of single crystal AlN without any complex fabrication, substrate removal, planarization or bonding.

**Ge in $SiO_2$**: The resonant antenna designs are patterned into a 500µm thick quartz substrate using a negative photoresist (AZ nLOF 5050) using stepper lithography and an ICP etch for the $SiO_2$ to form engraving of 1.5µm. 1µm Ge wires were deposited using an E-beam evaporation method and the excess Ge was lifted using an heated NMP solution at 80°C for 4 hours. The resonators were capped off with another conformal coating layer of PECVD $SiO_2$ of 500nm. This process enables us to embedd high index Ge inside the quartz substrate while maintaining a resaonable flat top surface at the end of the process.

**Intrinsic InSb in doped InSb:** Molecular Beam Epitaxy growth of the layer structure (undoped GaSb (001) substrate/thin GaSb buffer/ 1.2 µm Te-doped InSb/1 µm i-InSb) was performed using a modified VG V80H III-V MBE system with a base pressure <$2 \times 10^{-11}$ Torr. The undoped GaSb substrate was thermally desorbed under an $Sb_2$ overpressure, after which a thin buffer of unintentionally doped GaSb was grown. The substrate temperature was then lowered to less than 380°C for the growth of nominally $1\times10^{19}$/cm$^3$ Tellurium doped InSb at a growth rate of 0.56 μm/hour. This was then followed by a growth of undoped InSb at the same growth temperature and growth rate. After MBE growth, the sample was removed from vacuum and a 300nm thick film of SiO$_2$ was deposited using Plasma enhanced CVD to form a hard mask for subsequent dry etching. Photolithography was performed using a projection stepper aligner with SPR 955 as the photoresist (PR) along with Contrast Enhancing Mixture (CEM) top layer to increase verticality of the developed PR. The hard mask was etched using an inductively coupled plasma (ICP) dry etch using CHF$_3$/Ar gas mixture followed by a 10 minute oxygen plasma clean at 250°C to remove any organic residue from the sample surface. The InSb was etched using a Reactive Ion Etching (RIE) process using a methane plasma and timed to stop at an etch depth of 1 μm. The excess hard mask was removed using a wet etch in a buffered HF solution for 60s. The resonators were subjected to a high temperature (275°C) oxygen plasma etch for 8 minutes to remove any traces of carbon containing residues from the surface. The dry etched surface was prepared for the next MBE growth step using sequential etching in diluted HNO$_3$ 1:10 DI water solution and diluted HF 1:10 DI water solution. The sample was dipped in both solutions one after the other for 6s each for a total of one minute. (HNO$_3$) oxidizes the all exposed surfaces and (HF) preferentially etches the oxidized layers. The sample was then loaded into an interconnected UHV system, immediately after the surface preparation etch. Approximately six months elapsed between the first epi-layer growth and the reintroduction of the sample in the UHV system. The surface was then cleaned in a UHV preparation chamber (base pressure $<1\times10^{-10}$ Torr) with an MBE Komponenten atomic hydrogen source at a substrate temperature of 400°C and a chamber pressure of $5\times10^{-6}$ Torr for one hour. The sample was then transferred in-vacuo to the the modified VG V80H III-V MBE for the final epi-layer re-growth. A 2.7 μm thick, $2\times10^{-19}$/cm$^3$ Silicon doped n-InSb layer was then regrown on the hydrogen cleaned intrinsic InSb resonators at 370°C at a growth rate of 2 μm/hour. This regrowth was found to lead to a smooth surface a c(4x4) termination in Reflection High Energy Electron Diffraction (RHEED).. This process flow enables us to epitaxially regrow a doped InSb layer on a dry etched InSb substrate to form embedded resonators.

**FTIR Microscope measurements:** The temperature dependent and size-dependent scattering properties were measured in a Bruker v70 FTIR miscrope similar to the methods previously described in references [28–30,54] The Q-factor of the resonance based on the dip in the differential reflection plot shown in Fig 2B is based on the half-width half-maximum (HWHM) of the resonance due to overlapping modes at longer wavelengths (Fig 3B). The temperature dependent reflectivity measurements were normalized to Au, and done using a FTIR microscope on thin film region away from the resonators to get the signal from just the n-InSb/GaSb substrate.

**Acknowledgments**: This work was supported by the US Department of Energy (DOE), Office of Basic Energy Sciences, Division of Materials Sciences and Engineering and performed, in part, at the Center for Integrated Nanotechnologies, an Office of Science User Facility operated for the US DOE Office of Science. Sandia National Laboratories is a multi-mission laboratory managed and operated by National Technology and Engineering Solutions of Sandia, LLC, a wholly owned subsidiary of Honeywell International, Inc., for the US DOE's National Nuclear Security Administration under contract no. DE-NA0003525. This Article describes objective technical results and analysis. Any subjective views or opinions that might be expressed in the paper do not necessarily represent the views of the US DOE or the United States Government. This study was in part funded by the US DOE Basic Energy Science Program (BES20017574). J.A.S was supported by the Office of Naval Research (Grant #N00014–22–1–2337). The authors acknowledge the Solid-State Lighting and Energy Center (SSLEC) at UCSB. M.P. and C.J.P. would like to acknowledge support from the Vannevar Bush Faculty Fellowship program sponsored by the Basic Research Office of the Assistant Secretary of Defense for Research and Engineering and funded by the Office of Naval Research through grant N00014-15-1-2845. A portion of this work was performed in the UCSB nanofabrication facility, part of the National Science Foundation (NSF)-funded National Nanotechnology Infrastructure Network (NNIN).

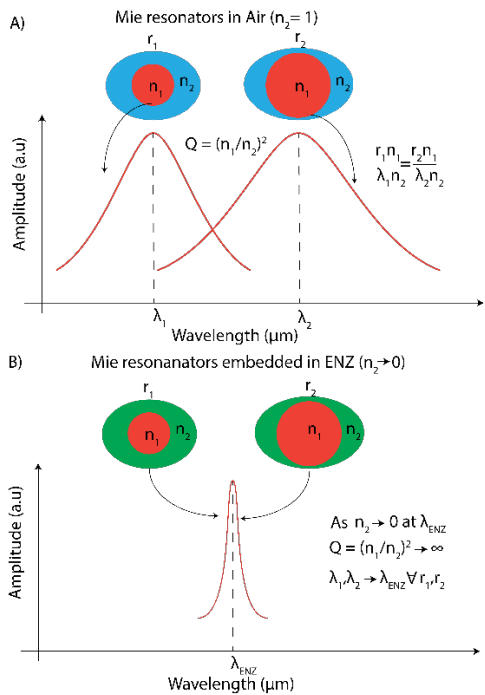

**Figure 1. Embedding Mie resonators in ENZ materials**: A) Schematic representing the geometric dispersion associated with the size of the resonator ($r_2 > r_1 \rightarrow \lambda_2 > \lambda_1$) and resonant wavelength when high-index resonators ($n_1 > n_2$) are scattering light in a lower index media (e.g. air $n_2 = 1$). Such that $\frac{n_1 r_1}{n_2 \lambda_1} = \frac{n_1 r_2}{n_2 \lambda_2}$ B) In the scenario, that $n_2 \rightarrow 0$ at $\lambda_{ENZ}$, the geometric dispersion associated with the size of the resonator vanishes and the resonant quality factor (Q) diverges to ∞, limited by the losses in $n_2$.

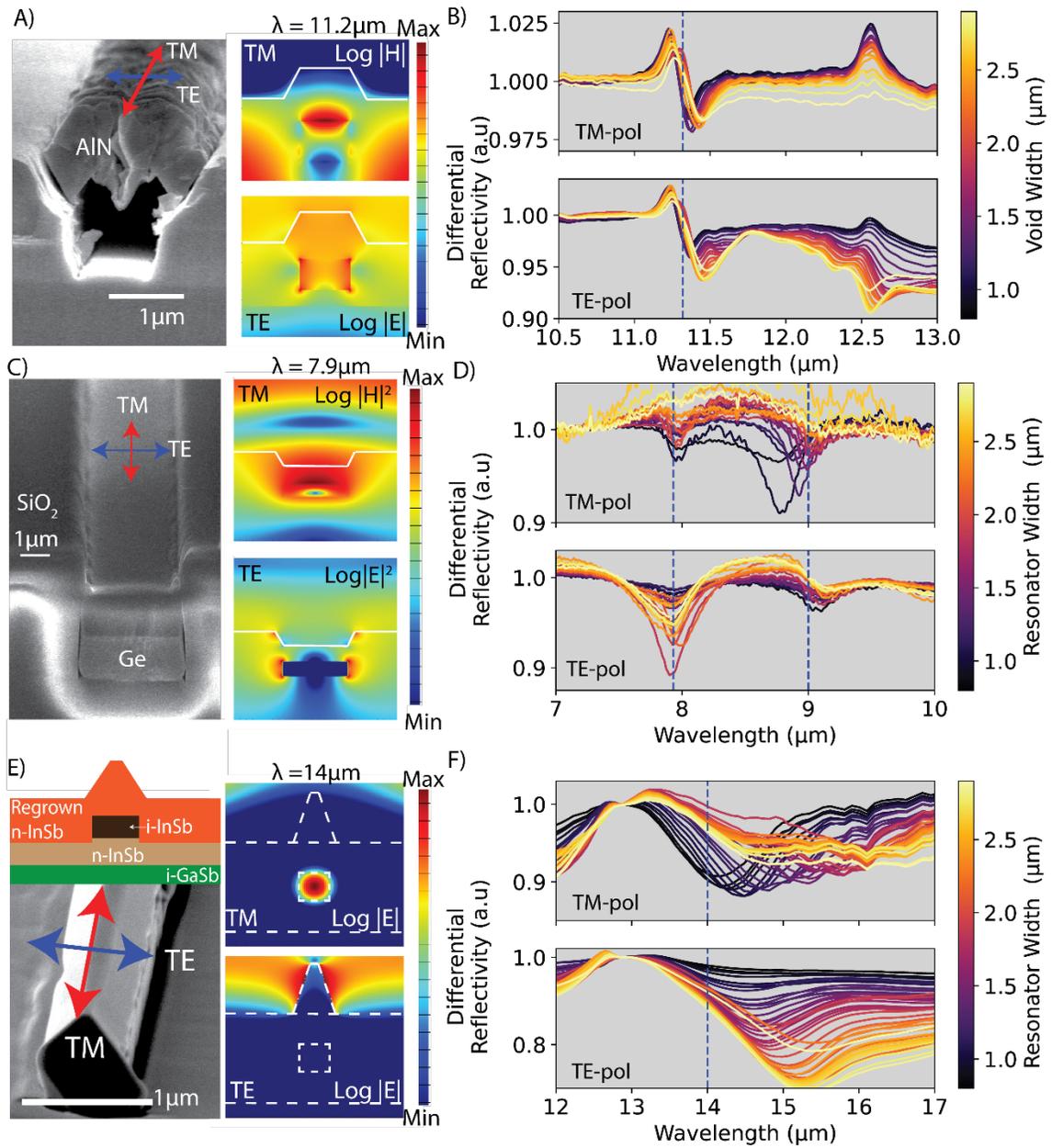

**Figure 2: Mie resonances at ENZ wavelength: A)** Scanning electron microscope image of the void resonator inside AlN thin film. The right side panel shows the electric and magnetic field distributions for the resonant electric dipole mode at the ENZ wavelength (11.2μm) of the AlN media. **B)** Plot shows the differential reflection (=1- Reflection of the resonator normalized to Au subtracted from the reflection of the background normalized to Au from the same area) of a single resonator (voids in AlN of varying widths as shown on the colorbar on the right) under TM (top panel) and TE (bottom panel) polarizations. The dips in the differential reflection curves represent the resonances and the strength of resonances quantified based on its deviation from 1. The blue dashed line shows the ENZ wavelength of AlN where the high Q-

factor (~95) resonances are formed. **C)** SEM image of the resonator cross section with Ge nanowire embedded inside SiO$_2$ medium on a quartz substrate. The TE and TM arrows indicate the polarization definitions with respect to the nanowire axis. The electric (TE, bottom) and magetic (TM, top) field intensity plots of the resonant field profiles at the ENZ wavelength of the oxide showing the excited electric and magnetic dipole resonances respectively. **D)** Similar to panel B for Ge antennas embedded within SiO$_2$. Plots showing the differential reflection spectra of the TM (top panel) and TE (bottom panel) for varying resonators widths as indicated in the colorbar on the right. The blue dashed line indicates the ENZ wavelength of the SiO$_2$. **E)** Plot showing the cross-section sketch and scanning electron microscopy image of the resonator of width 800nm and the resonant field profiles under TE and TM polarization. Note that only the TM wave excites the resonances in the embedded i-InSb resonator while the TE wave excites the regrown faceted ridge of n-InSb. **F)** Similar to panel B,D for i-InSb antennas embedded within n-InSb. Plots showing the differential reflection spectra of the TM (top panel) and TE (bottom panel) for a varying resonator widths as indicated on the right colorbar. The blue dashed line indicates the ENZ wavelength of the n-InSb at 14µm.

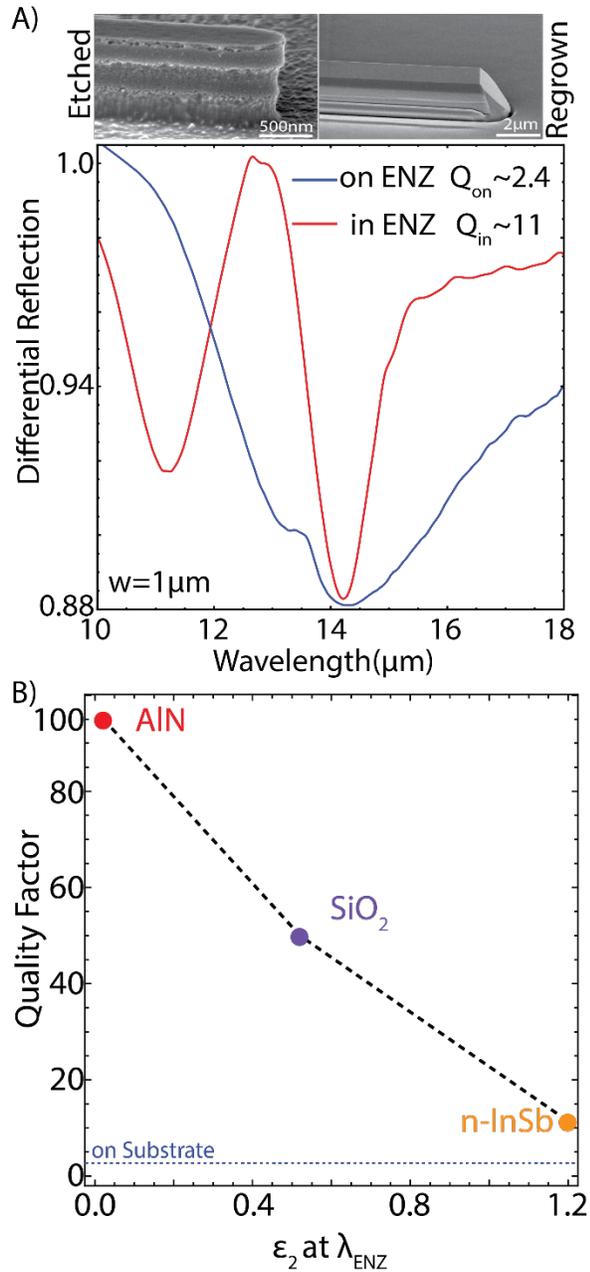

**Figure 3: Q-factor Comparison:** **A)** Scanning electron microscopy image of the dry etched InSb resonator on the ENZ substrate showing the vertical edge of the resonator and image of the regrown resonator showing the faceted ridge covering the vertically etched intrinsic InSb sidewalls. The reflection spectra of the resonator of width 1μm on the doped InSb substrate (blue) and inside the doped InSb cavity (red) showing the narrowing of the fundamental TM resonance. The normalized line-width Q-factor (=$\lambda_R/\delta\lambda_{FWHM}$) increases by a factor of 4.5 by embedding the resonator inside the ENZ cavity. **B)** The Q-factor of the resonances at the ENZ wavelength decreases as the absorption at the ENZ

wavelength increases. The plot compares three different material systems where the measured Q-factor can be increased from the on-substrate limit of 2.4 in high-index i-InSb resonant system to 95 for voids in AlN.

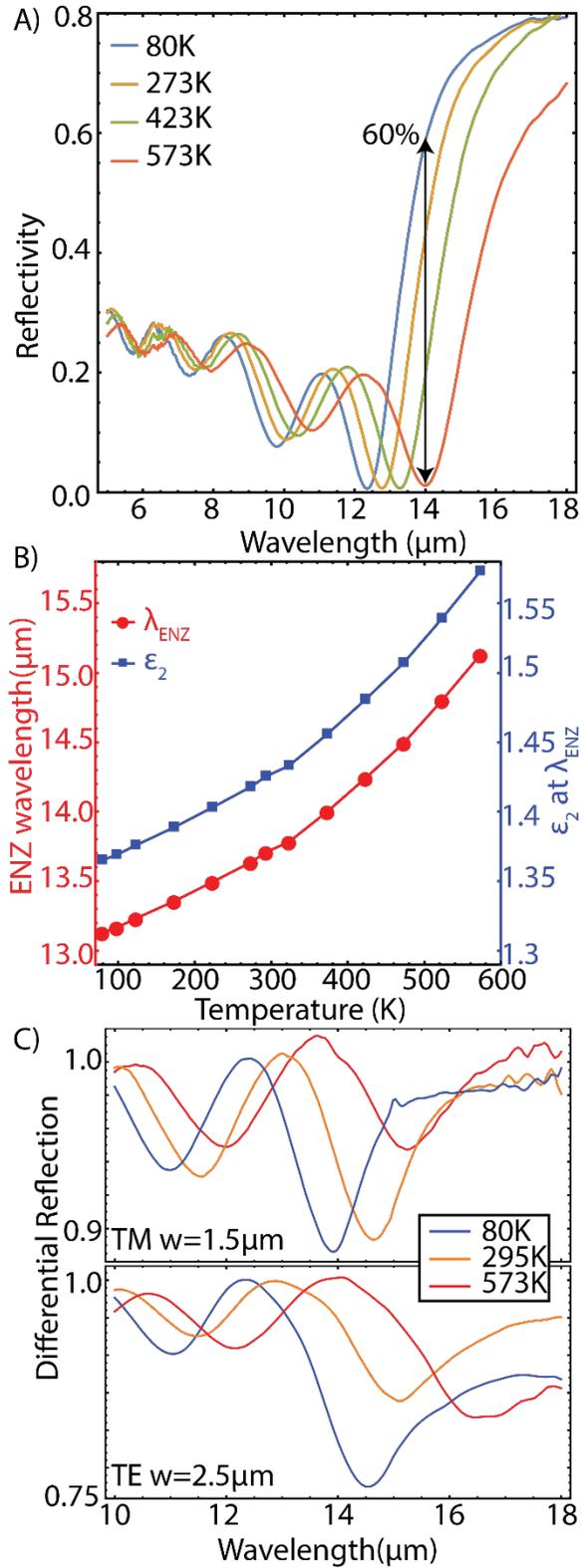

**Figure 4: Thermal Tunability: A)** Reflection of the doped InSb cavity surrounding the resonator in the mid-infrared wavelength at different temperatures from 80K to 573K. The black arrow indicates the continuous reflectivity change of 60% at 14µm due to shifting plasma edge with increase in temperature.

**B)** The thermal dispersion of ENZ wavelength ($\lambda_{ENZ}$ left axis, red) and $\varepsilon_2$ at $\lambda_{ENZ}$ (right axis, blue) in the doped InSb fitted to the analytical reflectivity curves using Transfer matrix model. We show that the $\lambda_{ENZ}$ of the doped InSb cavity can be continuously tuned with temperature over 2µm with absorption losses $\varepsilon_2$<1.6. **C)** Plots showing the thermal dispersion of the reflection spectra for TM (w=1.5µm, top panel) resonances and TE (w=2.5µm, bottom panel) resonances from 80K to 573K. The narrow TM resonance is shifted by 170% of the linewidth.

# Supplementary Information

# High-Q, size-independent, and reconfigurable optical antennas via zero-index material dispersion engineering

Prasad P. Iyer[1,4*], Mihir Pendharkar[1], Anchal Agrawal[1], Mike Iza[2,3], Stacia Keller[2,3], Umesh K. Mishra[1,3], Shuji Nakamura[1,2,3], Steven DenBaars[1,2,3], Chris Palmstrom[1,2] and Jon A. Schuller[1*]

1. Electrical and Computer Engineering Department, University of California Santa Barbara
2. Material Science and Engineering Department, University of California Santa Barbara
3. Solid State Lighting and Energy Center, University of California Santa Barbara
4. Center for Integrated Nanotechnologies, Sandia National Laboratory, Albuquerque, New Mexico

*ppadma@sandia.gov, jonschuller@ece.ucsb.edu

Table of Contents:

1. Thin film properties of ENZ materials
2. Geometric dispersion of resonators embedded in ENZ media

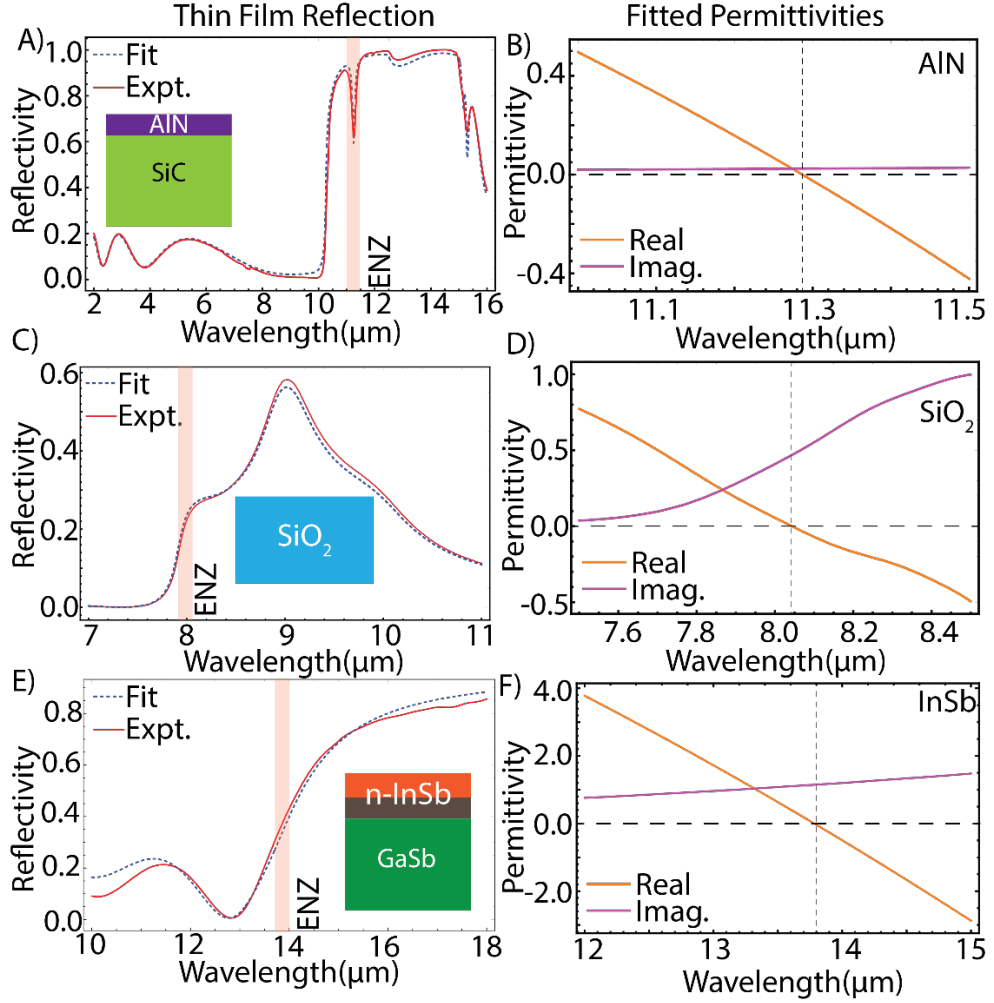

**Figure S1**: Panels A,C and E represent the refelction measurements in FTIR on thin films of the ENZ material systems (A: AlN, B: SiO₂ and C: n-InSb) in solid red and their corresponding fits in dashed blue lines. The orange band in each panel represents the ENZ region. Panels B,D, and F plotted the fitted real and imaginary parts of the dielectric permittivity of the ENZ materials from the fits shown in the left side

The permittivity for the polar nitrides GaN, AlN and SiC was modelled using equation 1:

$$\varepsilon(\omega, \varepsilon_\infty, \Omega_L, \Gamma, \Omega_T, n, \gamma, m_e) = \varepsilon_\infty \times$$

$$\left[1 + \left(\frac{(\omega_L^2 - \omega_T^2)}{(\omega_T^2 - \omega^2 - i\Gamma\omega)}\right) - \left(\frac{n \times e^2}{(\varepsilon_0 \times m_e)(\omega^2 + i\gamma\omega)}\right)\right] \quad (1)$$

Where: n: Electron density , e: Elementary charge (charge of an electron) , $\varepsilon_0$: Permittivity of free space, $m_e$: Electron mass, $\varepsilon_\infty$ : High-frequency permittivity of the material , $\omega_L$ : Longitudinal optical phonon frequency, $\omega_T$: Transverse optical phonon frequency, Γ: Damping constant for optical phonons, n: Electron density, γ: Electron damping coefficient, c: Speed of light

**Material Properties Table [1-8]:**

| Property | GaN | AlN | SiC |
|---|---|---|---|
| $\varepsilon_\infty$ | 5.29 | 4.61 | 6.7 |
| $m_e$ (relative to $m_o$) | 0.2 $m_o$ | 0.4 $m_o$ | 0.42 $m_o$ |
| $\omega_L$ (cm$^{-1}$) | 835 | 883.4 | 972 |
| $\omega_T$ (cm$^{-1}$) | 533 | 664 | 799 |
| Γ (cm$^{-1}$) | 10 | 2 | 5.1 |
| γ | e / (0.126 $m_e$) | e / (0.03 $m_e$) | e / (11 $m_e$) |

The permittivity model for the doped-InSb [9] and SiO$_2$ [10] was taken obtained from previously published literature.

S2: Geometric dispersion of the resonators embedded in ENZ media

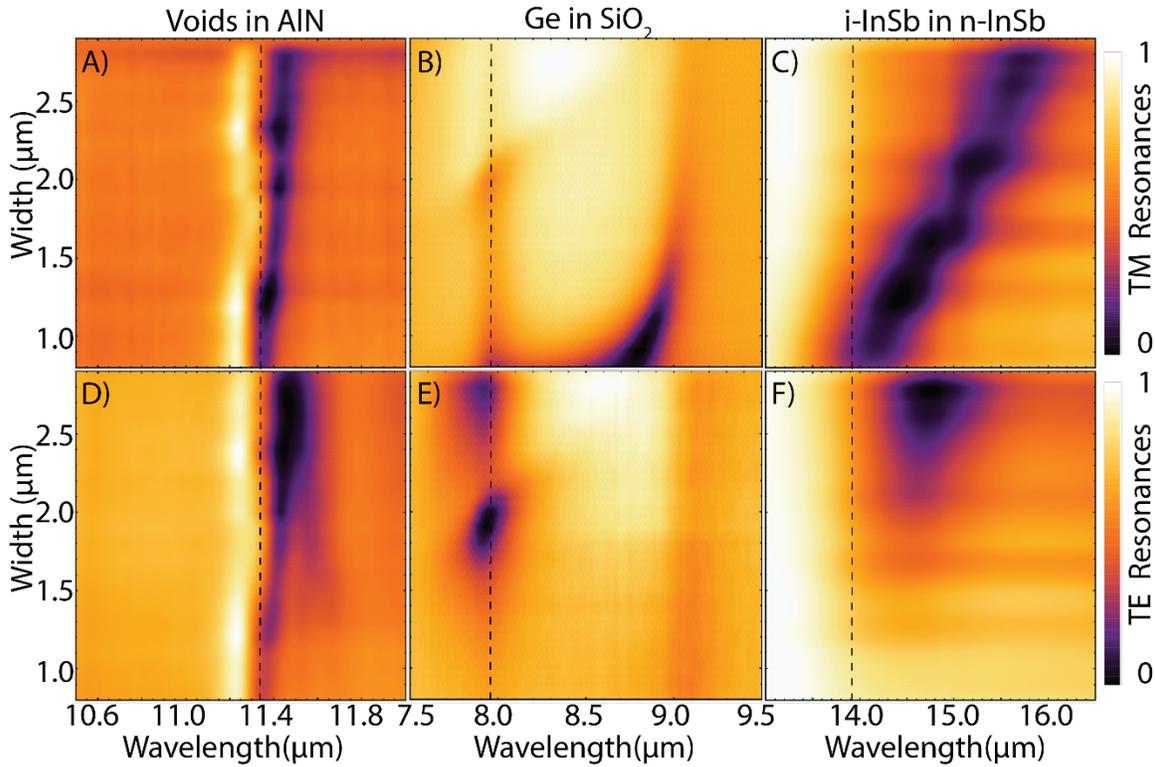

**Figure S2: Geometric Dispersion:** Normalized differential reflection color maps of the transverse magnetic (top panels: A, B, & C) and electric (bottom panels: D, E and F) resonances as a function of the width of the resonators (left axis) and wavelength (bottom axis). The plot shows the resonances as dips in reflection (dark regions). The veritcal dashed lines shows the ENZ wavelength for the material system at 11.32μm for AlN (A and D), 7.9μm in $SiO_2$ (B and E) and 14.0μm in n-InSb (C and F).